\documentclass{appolb}
\usepackage{epsfig}
\usepackage{amsmath,amssymb}
\usepackage{slashed}
\usepackage{color,array,tabularx}

\definecolor{holger}{rgb}{0.9,0,0}

\definecolor{michael}{rgb}{0.9,0,0.9}

\definecolor{stefan}{rgb}{0,0.6,0}

\newcommand{\Nf}{N_\mathrm{f}}

\begin{document}
\title{An asymptotic-safety mechanism for chiral Yukawa systems
}

\author{Michael M. Scherer\thanks{Presented by M.M.S.~at 49th Cracow School of Theoretical
Physics, June 2009, Zakopane}, Holger Gies, 
\and
Stefan Rechenberger
\address{Theoretisch-Physikalisches Institut, Friedrich-Schiller-Universit\"at, Max-Wien-Platz 1, D-07743 Jena, Germany}
}
\maketitle
\begin{abstract}
We introduce Weinberg's idea of asymptotic safety and pave the way
towards an asymptotically safe chiral Yukawa system with a
$U(N_\mathrm{L})_\mathrm{L}\otimes U(1)_\mathrm{R}$ symmetry in a
leading-order derivative expansion using nonperturbative functional RG
equations. As a toy model sharing important features with the standard model
we explicitely discuss $N_\mathrm{L}=10$ for which we find a non-Gau\ss ian
fixed point and compute its critical exponents. We observe a reduced hierarchy
problem as well as predictions for the toy Higgs and the toy top mass.
\end{abstract}

\section{Introduction}

Quantum field theory (QFT) has been very successful in the description of a
large number of phenomena, in particular in high energy physics. However there
is a widespread belief that QFT in many cases only accounts for effective
theories and that it is not suited to constitute a fundamental theory but
should be replaced by another concept at a microscopic scale. This is due to
the apparent non-renormalisability of important action functionals, e.g. the
Einstein-Hilbert action, describing gravity
\cite{'tHooft:1974bx,Goroff:1985sz,vandeVen:1991gw,Christensen:1979iy}. Furthermore
the standard model of particle physics is plagued by the problem of triviality
in the sector describing quantum electrodynamics (QED)
\cite{Landau,Gell-Mann:fq,Gockeler:1997dn,Gies:2004hy} and in the Higgs sector
\cite{Wilson:1973jj,Luscher:1987ek,Hasenfratz:1987eh,Heller:1992js,Callaway:1988ya,Rosten:2008ts},
which forbids to extend the standard model beyond a certain ultraviolet
scale. However the issue of non-renormalisability is often adressed within
\emph{perturbative} QFT, which is not an essential concept of QFT itself. For
perturbative QFT to apply it has to be possible to expand about vanishing
interactions, i.e., about the Gau\ss ian fixed-point (GFP), which is a severe
limitation. A more general conceptual understanding of QFT, not sticking to
the constraint of perturbativity, was elucidated by Steven Weinberg when he
introduced the idea of asymptotic safety
\cite{Weinberg:1976xy,Percacci:2007sz,Weinberg:2009}. In an asymptotically
safe QFT the microscopic action entering the functional integral approaches an
interacting fixed point in the space of action functionals in the infinite UV
cutoff limit. Thus no unwanted divergencies can occur. This renders the theory
well-defined on all scales. The scenario has already been applied to a number
of models ranging from four-fermion models
\cite{Rosenstein:pt,Gies:2003dp,Schwindt:2008gj}, simple Yukawa systems
\cite{GiesScherer:2009,Scherer:2009cy}, nonlinear sigma models in $d>2$
\cite{Codello:2008qq}, extra-dimensional gauge theories \cite{Gies:2003ic},
and gravity
\cite{Reuter:1996cp,Percacci:2003jz,Gies:2009gb,Saueressig:2009,Niedermaier:2006wt}.

The idea of asymtpotic safety has gained considerable attention in the context
of gravity, especially in the last 10 years. In fact, there is a lot of
evidence that such an interacting fixed point exists for diffeomorphism
invariant actions, allowing for a formulation of a non-perturbative
renormalisable quantum field theory of gravity
\cite{Reuter:1996cp,Percacci:2003jz,Gies:2009gb,Saueressig:2009}. The calculations necessary
to compute the RG flow in theory space, however, are tedious and involve a
number of techniques, such as e.g. the background field method, that
complicate the discussion of the bare concept of asymptotic safety
considerably. It would therefore be convenient to have a simpler setting,
where the idea and the basic concepts of asymptotic safety could be tested and
understood in a transparent manner.

On the quest for such a simpler setting the standard model with its problem of
triviality might be a beacon. Here, the Landau poles of perturbation theory in
the QED and the Higgs sector suggest that one should introduce new degrees of
freedom for a fundamental description of particle physics. However, before
doing so a non-perturbative computation of the Higgs sector including fermions
and also gauge fields would have to show whether the problem of triviality
still persists or whether the theory might be asymptotically safe in
Weinberg's sence, by acquiring a fixed point in the ultraviolet. As a step
towards this scenario, we investigate a toy model for the standard model
without gauge fields and with a particular chiral left/right asymmetry
\cite{Scherer:2009cy}. To leading order in the derivative expansion, we 
find an asymptotically safe theory which is well-defined on all scales
(renormalisable) with a highly predictive power and comparatively simple from
a computational point of view. Apart from a possible application to the
complete standard model of particle physics including a prediction of the
Higgs and the top mass, it also allows for a better understanding of how
asymptotic safety works.

This contribution is organised as follows: in section \ref{sec:problems} we
briefly recall triviality and the hierarchy problem as they occur in the Higgs
sector of the standard model. Section \ref{sec:as} introduces the idea of
asymptotic safety and the Wetterich equation, which is our tool to investigate
QFT non-perturbatively. In section \ref{sec:yukawa} we discuss our toy model,
a particular chiral Yukawa system, whose fixed-point structure is analysed in
section \ref{sec:fixedpoints}. Conclusions are drawn in the last section.

\section{Two problems of the Higgs sector}
\label{sec:problems}

In the Higgs sector of the standard model we find the problem of triviality
and the hierarchy problem. To understand the nature of the triviality problem,
it suffices to consider a purely bosonic theory, where we mimic the
Higgs field in terms of a single component real scalar field $\phi$, with a
Lagrangian
\begin{equation}
 \mathcal{L}= \frac{1}{2}(\partial_{\mu}\phi)^2+\frac{{m}^2}{2}\phi^2
+\frac{{\lambda}}{8}\phi^4.
\end{equation}
Using one-loop RG-improved perturbation theory we can establish a relation
between the bare and the renormalized four-Higgs-boson coupling
$\lambda\phi^4$
\begin{equation}
 \frac{1}{\lambda_{\rm R}}-\frac{1}{\lambda_{\Lambda}}=\beta_0\ \mbox{Log}\left(
   \frac{\Lambda}{m_{\rm R}}\right),\ \beta_0=\rm{const.}>0,
\end{equation}
where $\lambda_{\Lambda}$ and $\lambda_{\rm R}$ are the bare and the
renormalised couplings, respectively, $\Lambda$ is the ultraviolet cutoff
scale and $m_{\rm R}$ the renormalised mass. For fixed non-zero (non-trivial)
$\lambda_{\rm R}$ and $m_{\rm R}$ the bare coupling $\lambda_{\Lambda}$ runs
into a pole at a finite UV scale, the Landau pole $\Lambda_{\rm L}$. This
indicates the breakdown of the perturbative QFT at this scale.

For a purely bosonic theory also a nonperturbative treatment of QFT (e.g., on
the lattice) confirms triviality, inhibiting the theory to be
fundamental. However the standard model Higgs sector of course contains
bosonic and fermionic degrees of freedom; as we show in
Sect.~\ref{sec:fixedpoints}, a balancing of the fermionic and bosonic quantum
fluctuations can make the pole disappear and render the system well-defined on
all scales.

Whereas the triviality problem is a conceptual problem, the hierarchy problem
is only a problem of the unnaturalness of strongly fine-tuned initial
conditions. We observe a huge hierarchy in the standard model between the
electroweak scale $\Lambda_{\rm EW} \sim 10^2 {\rm GeV}$ and e.g. the scale of
a grand unified theory $\Lambda_{\rm GUT}\sim 10^{16} {\rm GeV}$. The Higgs
mass renormalises quadratically with the UV cutoff $\Lambda$ and in
perturbation theory the relation between bare and renormalised mass is given
by
\begin{equation}
 m_{\rm R}^2 \sim m_{\Lambda, \rm UV}^2 -\delta m^2.
\end{equation}
Here, the renormalised mass is of order $m_{\rm R}^2 \sim 10^4 {\rm GeV}^2$,
and the fluctuation contribution is of the order of the cutoff $\Lambda^2\sim
\delta m^2 = X \cdot 10^{32}\rm{GeV}^2$ (here, $X$ is a pure number depending
on the coupling values). As a consequence, the counterterm has to be
$m_{\Lambda, \rm UV}^2 = 10^{32}(X+... 10^{-28}) {\rm GeV}^2$. It is this
second term in parentheses which has to be fine-tuned to a relative precision
of $\Lambda_{\rm EW}^2/\Lambda_{\rm GUT}^2 \sim 10^{-28}$. In an RG language
this quadratical cutoff dependence corresponds to a renormalisation at a
noninteracting (Gau\ss ian) fixed point with a critical exponent $\Theta
=2$. A nonperturbative analysis with an interacting fixed point might have small
critical exponents, e.g. close to zero, which could lead to a weaker cutoff
dependence and the chance to make the hierarchy problem disappear. We present
such a computation with a reduced hierarchy problem in Sect.~\ref{sec:fixedpoints}.

\section{Asymptotic safety and the flow equation}
\label{sec:as}

\begin{figure}
\begin{center}
\includegraphics[width=.49\textwidth]{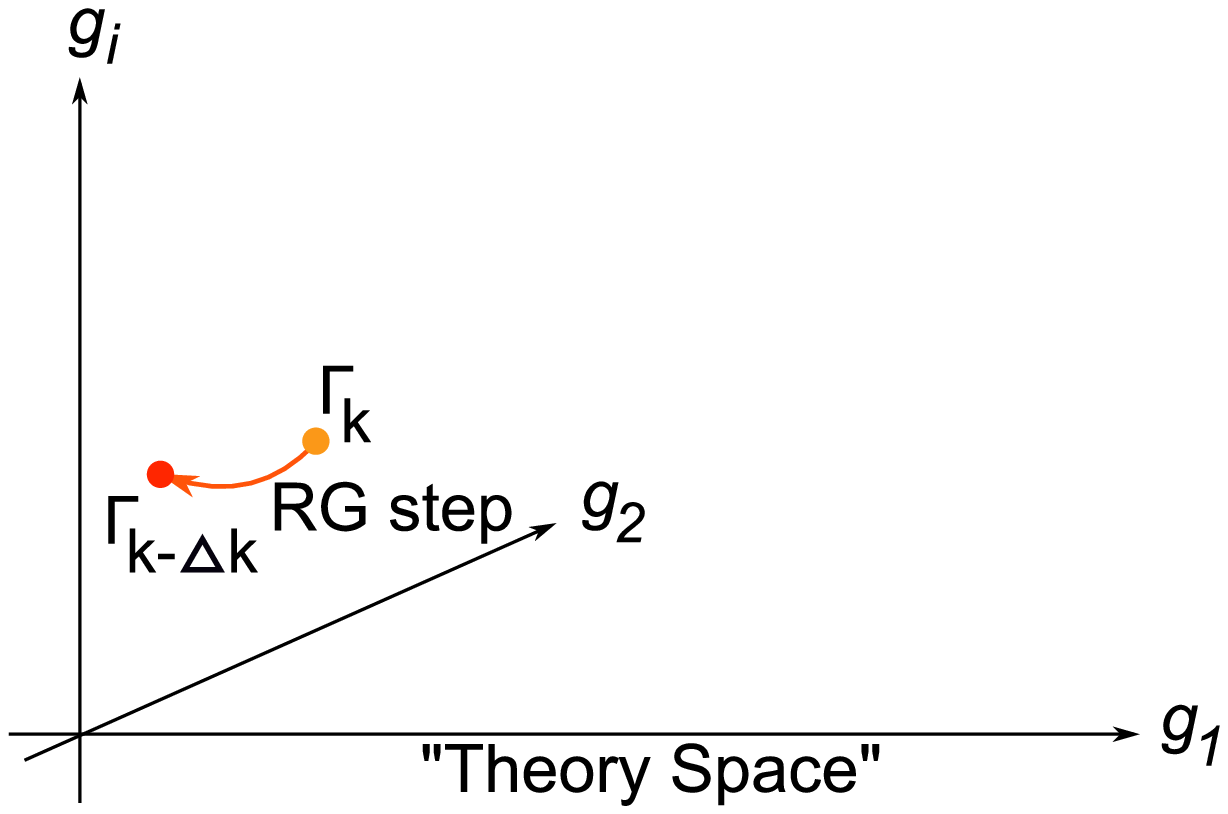}
\includegraphics[width=.49\textwidth]{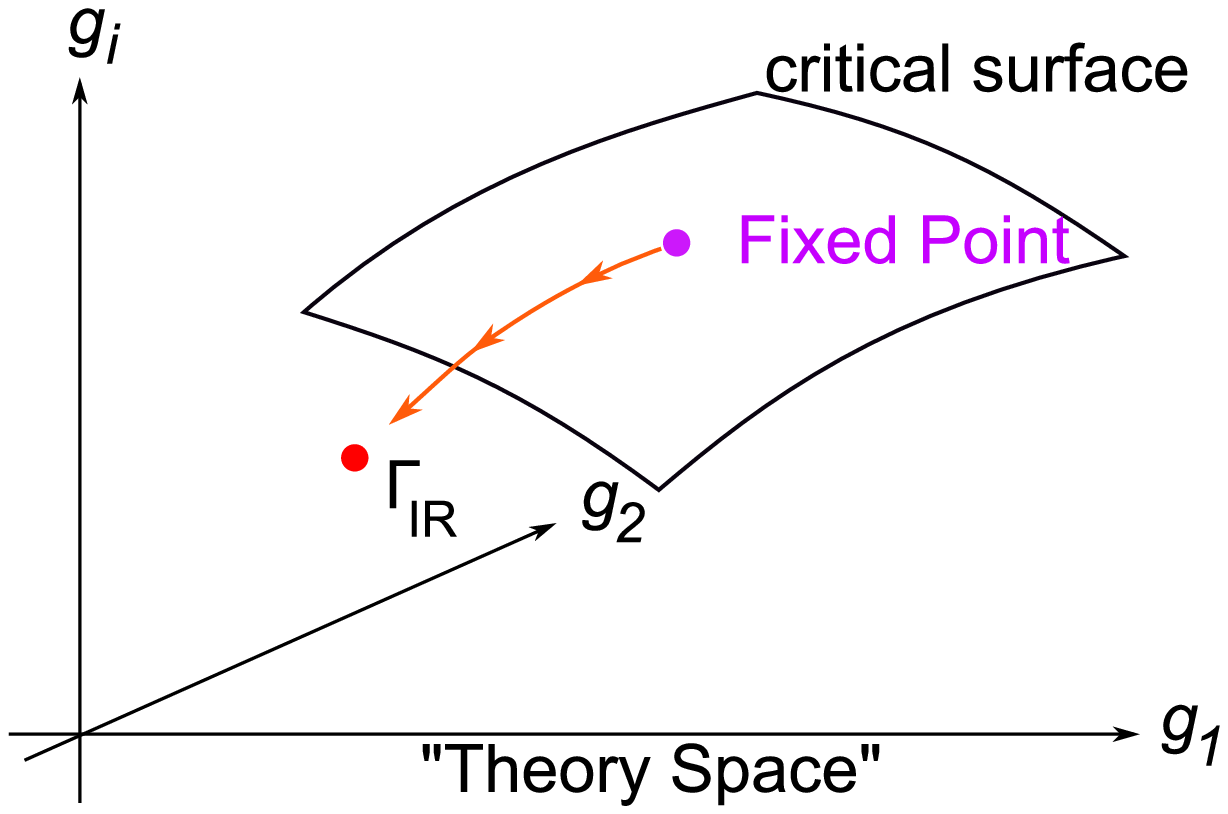}
\end{center}
\caption{Sketch of a 3-dimensional subspace of theory space spanned by three
  operators with associated couplings $g_1,g_2$ and $g_i$. Left panel: an RG
  step shifts the effective average action at a scale $k$ to a different
  point at a lower scale. Right panel: RG flow from an ultraviolet
  fixed point to a physical infrared effective action.}
\label{fig:theoryspace03}
\end{figure}

In this section we sketch the idea of asymptotic safety, comprehensive reviews
on asymptotic safety can be found in
\cite{Percacci:2007sz,Niedermaier:2006wt}. Consider an effective average
action $\Gamma_k [\chi]$ of an effective quantum field theory at scale $k$.
This is an action functional in the set of fields $\chi$ and consists of
operators which are compatible with the underlying symmetries. $\Gamma_k
[\chi]$ contains all the fluctuations of the quantum fields with momenta
larger than $k$. It can be understood as an effective theory where a tree
level evaluation suffices to describe physics at scale $k$. We can think of
$\Gamma_k [\chi]$ as an expansion in terms of (dimensionless) running
couplings $g_{i,k}$ and all possible field operators ${\cal{O}}_i$.
\begin{equation}
 \Gamma_k[\chi]=\sum_i g_{i,k} {\cal{O}}_i,\ \mathrm{ e.g. }\ {\cal{O}}_i =
 \left\{ \chi^2, \chi^4, (\partial \chi)^2 , \dots\right\} \,.
\end{equation}
The dependence of the effective action on the scale $k$, i.e. the
renormalization group flow, is given by the $\beta$ functions of the running
couplings,
\begin{eqnarray}
\partial_t \Gamma_k[\chi]= \sum_i \beta_{i,k} {\cal{O}}_i\,, \mathrm{ where}\ \beta_{i,k} = \partial_t g_i\ \mathrm{ and}\  \partial_t=k\frac{d}{dk} \,.
 \end{eqnarray}
 The field operators span the theory space, as is shown in Fig.~\ref{fig:theoryspace03}.
On the left hand side of Fig.~\ref{fig:theoryspace03} a sketch of the RG flow
is given. The position of the effective average action $\Gamma_k [\chi]$ in
theory space is given by a set of coordinates of running couplings $\left\{
g_{i,k} \right\}$. As we lower the scale from $k$ to $k-\Delta k$ by an RG
step, the transformation of the running couplings and so the change of
position in theory space is described by the $\beta$ functions and we end up
at a different effective average action $\Gamma_{k-\Delta k} [\chi]$ at the
new scale.

Suppose there is a (possibly non-Gau\ss ian) fixed point in theory space (see
r.h.s. of figure \ref{fig:theoryspace03}) where $\beta_{i,k} =0\ \forall \ i$.
If we can find an RG trajectory which connects the fixed point with a
meaningful physical theory represented by an effective average action
$\Gamma_{\mathrm{IR}}$ at some infrared scale, then we have found a quantum
field theory, which can be extended to arbitrarily high scales, since for the
cutoff scale $\Lambda \rightarrow \infty$ we just run into the fixed point and
no pathological divergencies can appear. This solves the triviality
problem. Note that in the perturbative setting of the standard model, the only
fixed point is the Gau\ss ian fixed point, which is not connected to a
physically sensible (non-trivial) effective action in the IR.

\begin{figure}
\begin{center}
\includegraphics[width=.49\textwidth]{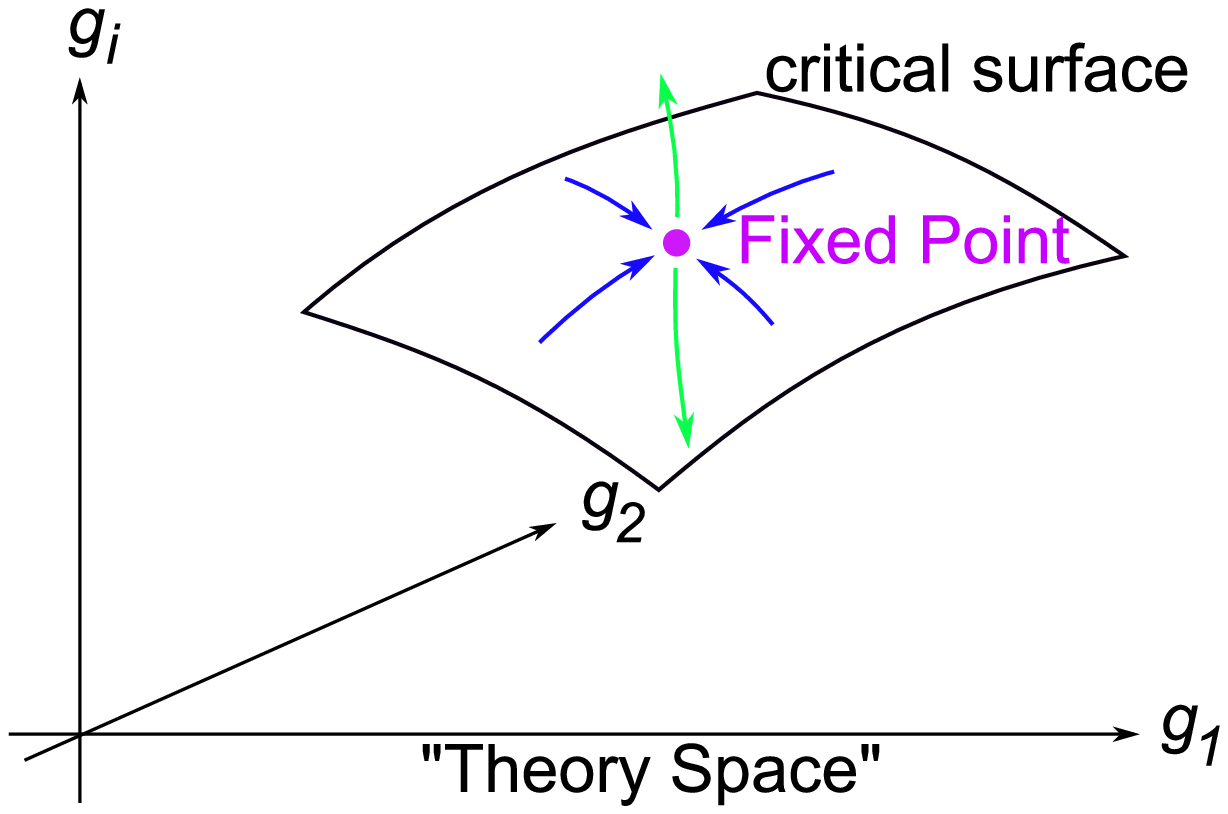}
\includegraphics[width=.49\textwidth]{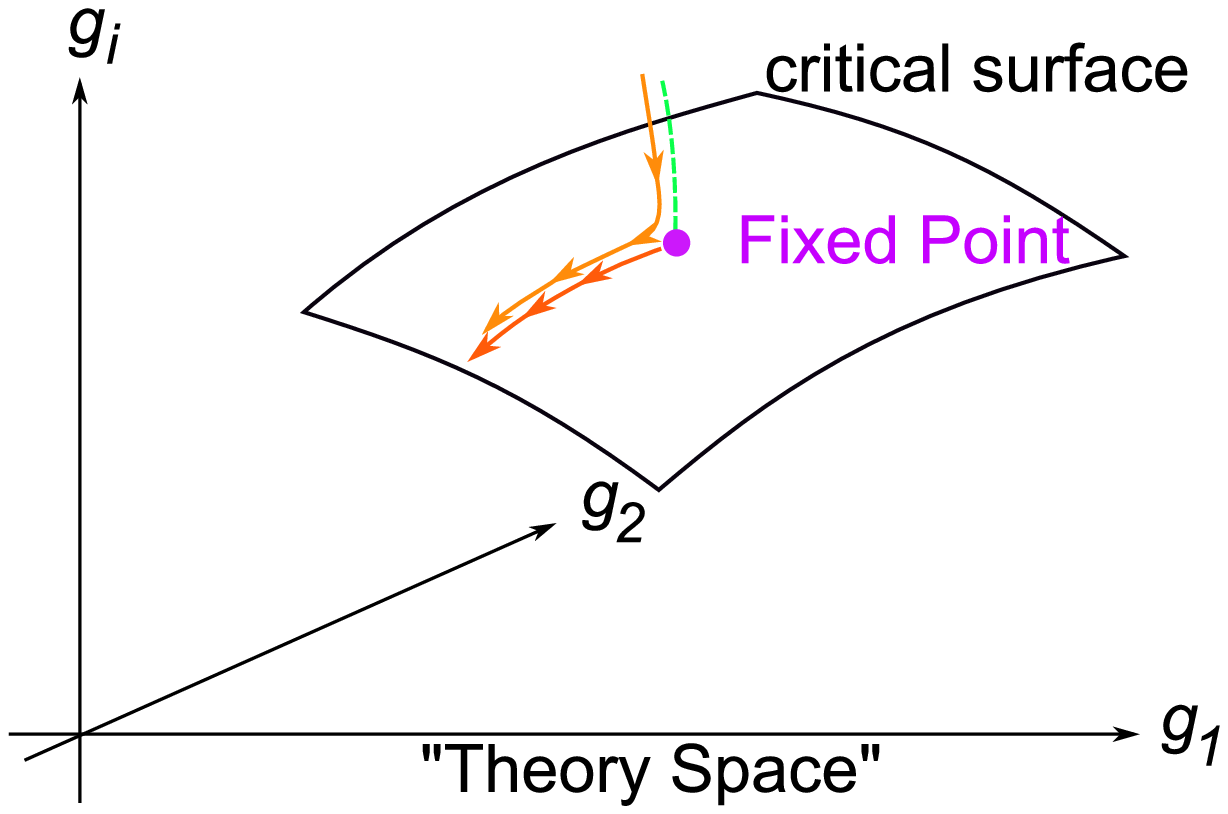}
\end{center}
\caption{Left panel: flow towards the UV in the vicinity of the fixed point.
  Relevant directions approach the fixed point (blue arrows) and span the
  critical surface; irrelevant directions run away from the fixed point for
  $k\to \infty$ (green arrows). Right panel: flow towards the IR. As the
  irrelevant directions are strongly attracted to the critical surface, the IR
  observables of the theory are solely determined by the relevant
  directions. }
\label{fig:theoryspace04}
\end{figure}

In the vicinity of a fixed point $g^\ast = \{g_i^\ast\}$ we can study the
behaviour of the RG trajectories near the fixed point, using the linearized
flow equations
\begin{equation}
 \partial_t g_i=B_i{}^j(g_j-g_j^\ast),\ B_i{}^j=\frac{\partial
   \beta_i}{\partial g_j}\Big|_{g^\ast} + \mathcal{O}((g-g^\ast)^2).
\end{equation}
The solution reads
\begin{equation}
 g_i = g_i^\ast + \sum_I C_I V_i^I \left( \frac{k_0}{k}\right)^{\Theta_I},
\end{equation}
where the integration parameters $C^I$ define the initial conditions at a
reference scale $k_0$. Furthermore, the eigenvectors $V^I$ and the negative of
the eigenvalues $\Theta^I$ of the stability matrix $B_i^j$ satisfy $B_i{}^j
V_j^I=-\Theta^I V_i^I$. For the flow towards the UV, the directions in theory
space with $\mathrm{Re} \{ \Theta_I \} > 0$ (relevant directions) are
attracted  towards the fixed point, see Fig.~\ref{fig:theoryspace04}, left panel.
The set of trajectories which run into the fixed point is called the critical
surface S. The number of linearly independent relevant directions at the fixed
point corresponds to the dimension of S. These directions determine the
physics in the infrared and the number of physical parameters to be fixed. The
theory is predictive if dim(S) is finite. The directions with $\mathrm{Re} \{
\Theta_I \} < 0$ run away from the fixed point as we increase $k$, see
Fig.~\ref{fig:theoryspace04}, left panel. By contrast, as we decrease
$k$, the irrelevant directions rapidly approach the critical surface, as
displayed in Fig.~\ref{fig:theoryspace04}, right panel. Therefore, the
observables in the IR are all dominated by the properties of the fixed point,
independently of whether the flow has started exactly in or near the critical
surface.  This establishes the predictive power of the asymptotic safety
scenario.

If a critical exponent is much larger than zero, say of $\mathcal O(1)$, the
RG trajectory rapidly leaves the fixed-point regime towards the IR. Therefore,
separating a typical UV scale where the system is close to the fixed point
from the IR scales where, e.g., physical masses are generated requires a
significant fine-tuning of the initial conditions. In the context of the
standard model, the size of the largest $\Theta^I$ is a quantitative measure
of the hierarchy problem.

As the non-perturbative tool to search for a NGFP in the space of action
functionals and to compute the properties at this FP, we use the Wetterich
equation which provides a vector field $ \beta$ in theory space in terms of RG
$\beta$ functions. The flow of the effective average action $\Gamma_k$ is
determined by \cite{Wetterich:1993yh}:
\begin{equation}\label{flowequation}
	\partial_t\Gamma_k[\chi]
        =\frac{1}{2}\mathrm{STr}\{[\Gamma^{(2)}_k[\chi]+R_k]^{-1}(\partial_tR_k)\}.
\end{equation}
Here, $\Gamma^{(2)}_k$ is the second functional derivative with respect to the
field $\chi$. The function $R_k$ denotes a momentum-dependent regulator that
suppresses IR modes below a momentum scale $k$. The solution to the Wetterich equation provides for an RG trajectory
in theory space, interpolating between the bare action $S_\Lambda$ to be
quantized $\Gamma_{k\to\Lambda}\to S_\Lambda$ and the full quantum effective
action $\Gamma=\Gamma_{k\to 0}$, being the generating functional of 1PI
correlation functions; for reviews, see \cite{ReviewRG}.

\section{A chiral Yukawa system}
\label{sec:yukawa}

As a toy model for the standard model we employ a Yukawa theory with chiral fermions including one right-handed fermion $\psi_\mathrm{R}$ and $N_\mathrm{L}$ left-handed fermions $\psi_\mathrm{L}^a$, which are coupled to $N_\mathrm{L}$ complex bosons $\phi^a$ via a simple Yukawa interaction term $\bar h_k$ \cite{Scherer:2009cy}. The theory space is truncated by an action functional in \emph{leading-order} derivative expansion and reads
\begin{eqnarray}\label{eq:SULtruncation}
\Gamma_k&=&\int d^dx\Big\{i(\bar{\psi}_L^a\slashed{\partial}\psi_L^a+\bar{\psi}_R\slashed{\partial}\psi_R)+(\partial_{\mu}\phi^{a\dagger})(\partial^{\mu}\phi^a)\nonumber\\
&&\hspace{1.0cm}+U_k(\rho)+\bar h_k\bar{\psi}_R\phi^{a\dagger}\psi_L^a-\bar h_k\bar{\psi}_L^a\phi^{a}\psi_R\Big\}\,,
\end{eqnarray}
where we define $\rho=\phi^{a\dagger}\phi^a$. The index $k$ at the Yukawa
coupling and the effective potential shall indicate their scale dependence,
which will be governed by the Wetterich equation. The action is invariant
under chiral $U(N_\mathrm{L})_\mathrm{L}\otimes U(1)_\mathrm{R}$
transformations. Additionally to the \emph{perturbative} complications of
triviality and the hierarchy problem, this toy model also shares the feature
of a left-handed chiral sector with the Higgs-top sector of the standard
model. The number of left-handed fermions $N_\mathrm{L}$ is left as a free
parameter in order to study the dependence of a potential fixed point on a
varying number of degrees of freedom. Various models with Yukawa-type
interactions have already been studied within this derivative expansion
technique and yielded reliable results in low-energy QCD \cite{Jungnickel:1995fp},
critical phenomena \cite{Rosa:2000ju}, and ultra-cold fermionic atom gases
\cite{Birse:2004ha}.

\begin{figure}
\begin{center}
\includegraphics[width=.5\textwidth]{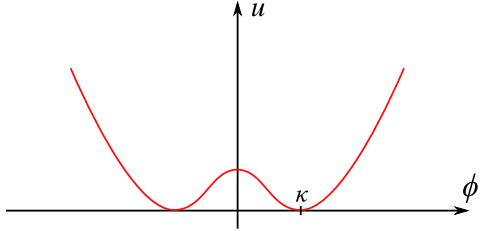}
\end{center}
\caption{Sketch of the scale-dependent effective potential $u$ as a function of the field $\phi_a$. For a theory in the regime of spontaneously broken symmetry  we employ an expansion about a positive minimum at $\kappa$, corresponding to a nonvanishing vacuum expectation value (vev) for the bosonic field.}
\label{fig:ssbpot}
\end{figure}

For the analysis of the fixed-point structure we introduce dimensionless quantities
\begin{eqnarray}\label{eq:dimensionless}
\tilde{\rho}&=&k^{2-d}\rho,\ \ h^2=k^{d-4}\bar h_k^2 ,\ \
 u(\tilde\rho)=k^{-d}U_k(\rho)|_{\rho=k^{d-2}\tilde\rho}.
\end{eqnarray}
The dimensionless effective potential $u$ is expanded about its dimensionless
minimum $\kappa:=\tilde\rho_{\mbox{min}}> 0$,
\begin{eqnarray}\label{eq:symeffpot}
  u&=&\frac{\lambda_{2}}{2!}(\tilde{\rho}-\kappa)^2+\frac{\lambda_{3}}{3!}(\tilde{\rho}-\kappa)^3+...\ \mathrm{with} \ \ \kappa,\ \lambda_{n_{\mathrm{max}}},\, \lambda_2>0.
\end{eqnarray}
This potential describes a theory in the regime of spontaneously broken
symmetry (SSB), which is depicted in figure \ref{fig:ssbpot}. The constraints
formulated in the expansion of the effective potential make sure that it is
bounded from below and constitutes an expansion about a positive minimum.
We could also think of an expansion about vanishing minimum, which would
describe the theory in the symmetric regime. However, the existence of a
suitable FP in the symmetric regime has been ruled out within the validity
limits of the derivative expansion \cite{Scherer:2009cy}.

\section{Fixed points and critical exponents}
\label{sec:fixedpoints}

\begin{figure}
\begin{center}
\includegraphics[width=0.65\textwidth]{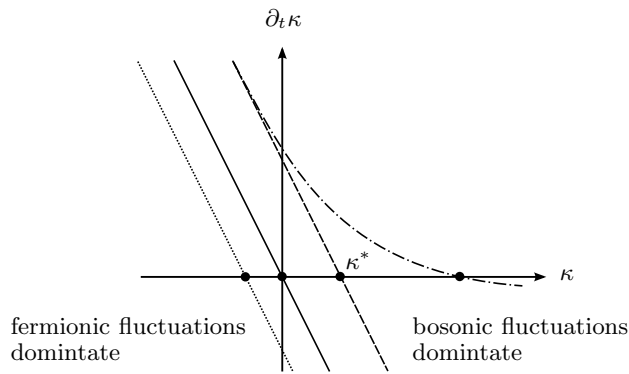}
\end{center}
\caption{Sketch of the $\beta$-function for the dimensionless squared Higgs vacuum expectation value $\kappa$. Dominating fluctuations of the boson field allow for a positive $\kappa^*$ (dashed line) and a suitable $\kappa$-dependence flattens the $\beta$-function near the fixed-point, which reduces the hierachy problem (dot-dashed line).}
\label{fig:sketch}
\end{figure}

To understand the occurence of fixed points in this model we investigate the loop contributions to the running of the dimensionless version of the squared bosonic field expectation value $\kappa$ in its flow equation of the form
\begin{equation}
\partial_t \kappa  = -2 \kappa + \mathrm{bosonic\ interactions}- \mathrm{fermionic\ interactions}.
\end{equation}
A positive sum of the contribution from the interaction terms gives rise to a
fixed point at $\kappa>0$ and allows for asymptotic safety, as we demonstrate
below. For a negative sum, no fixed point is possible
\cite{Scherer:2009cy}. Since fermions and bosons contribute with opposite
signs to the interaction terms, the existence of a fixed point $\kappa^\ast>0$
crucially depends on the relative strength between bosonic and fermionic
fluctuations, as is sketched in Fig.~\ref{fig:sketch}. In this figure the
solid line depicts the free massless theory with a trivial Gau\ss ian fixed
point at $\kappa=0$. If the fermions dominate, the interaction terms are
negative and the fixed point is shifted to negative values (being irrelevant
for physics), cf. dotted line. If the bosonic fluctuations dominate, the
$\kappa$ flow develops a non-Gau\ss ian fixed point at positive values
$\kappa^\ast>0$, cf. dashed line. This fixed point is UV attractive, implying
that the vev is a relevant operator near the fixed point. If the interaction
terms are approximately $\kappa$ independent, the slope of $\partial_t\kappa$
near the fixed point is still close to $-2$, corresponding to a critical
exponent $\Theta\simeq 2$ and a persistent hierarchy problem. An improvement
of ``naturalness'' could arise from a suitable $\kappa$ dependence of the
interaction terms that results in a flattening of the $\kappa$ flow near the
fixed point, cf. dot-dashed line.

The fixed point in the SSB regime induces a new mechanism for asymptotic
safety: near the fixed point, the vev exhibits a conformal behavior, being
always proportional to the actual RG scale $k$. If the vev was proportional to
a fixed threshold scale (as is usually the case during a symmetry-breaking
transition), the vev would induce the decoupling of massive modes below this
threshold. But as the vev runs with the scale, the system is always in the
onset domain of these threshold effects without showing any decoupling. In
other words, this conformal-vev mechanism renders the threshold effects strong
enough to induce the fixed point, but at the same time weak enough in order to
avoid decoupling.

\begin{figure}
\begin{center}
\includegraphics[width=0.60\textwidth]{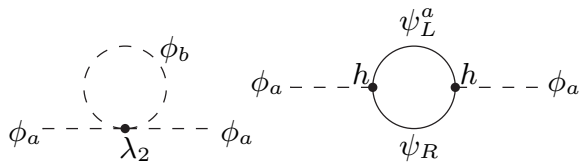}
\end{center}
\caption{Loop contributions to the renormalization flow of the Higgs dimensionless vev. The loop on the l.h.s. couples all available boson components giving a linear dependence on $N_\mathrm{L}$. This is not the case for the fermionic loop on the r.h.s. which is fully determined by the incoming boson field component.}
\label{fig:loops}
\end{figure}

For the conformal-vev mechanism to appear, the bosonic and fermionic
contributions have to be balanced. Whether or not this is possible depends on
the degrees of freedom and the algebraic structure of the theory: in
Fig.~\ref{fig:loops} we show the loop contributions from the bosonic and
fermionic fluctuations, which are particular for our model. The left loop
involves only inner boson lines. The vertex $\lambda_2$ allows for a coupling
between all available boson components. This implies a linear dependence on
$N_\mathrm{L}$ for the renormalization of the boson contribution. For the
fermion loop on the r.h.s of figure \ref{fig:loops} the incoming boson
$\phi_a$ fully determines the structure and does not allow for other
left-handed inner fermions than $\psi_{\mathrm{L}}^a$, inhibiting an
$N_\mathrm{L}$ dependence of this loop.

\begin{figure}
\begin{center}
 \includegraphics[width=0.32\textwidth]{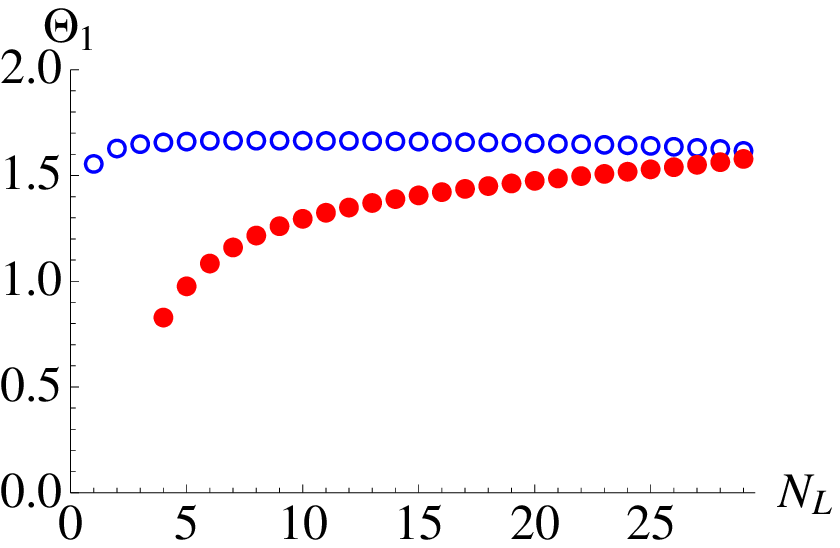}
 \includegraphics[width=0.32\textwidth]{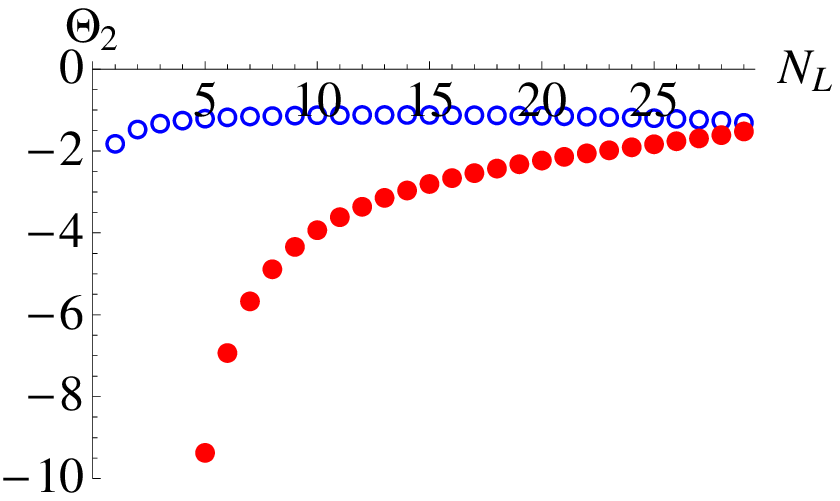}
 \includegraphics[width=0.32\textwidth]{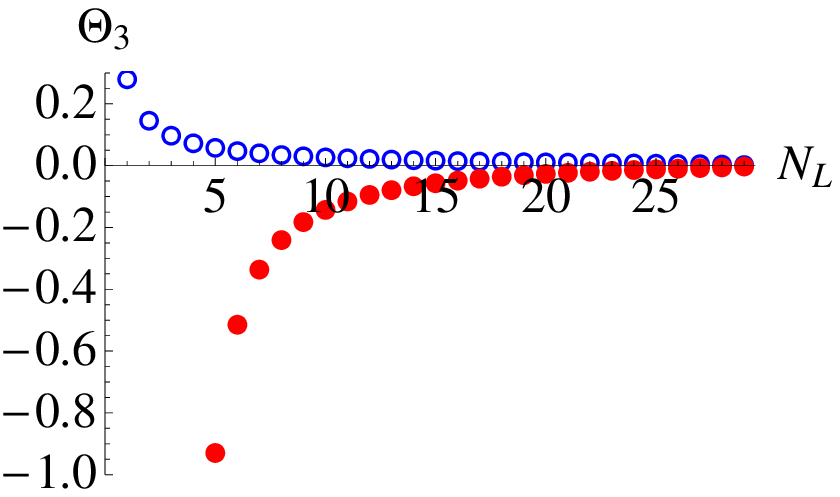}
\end{center}
\caption{Critical exponents for the NGFPs in the leading order truncation as a function of $N_\mathrm{L}$. The fixed point corresponding to the open circles 
has two relevant directions, whereas the fixed point corresponding to the filled circles has only one relevant direction.}
\label{fig:critexp}
\end{figure}

For a systematic analysis of the fixed-point structure let us start with a
very basic truncation only involving the flowing minimum of the effective
potential $\kappa$, the four-boson interaction $\lambda_2$ and the Yukawa
coupling $h$. For the fixed points we have to solve a set of nonlinear
algebraic equations of the form
\begin{eqnarray}
 \partial_t h^2 &=& \beta_h=0,\\
 \partial_t \lambda_2 &=& \beta_\lambda=0,\\
  \partial_t \kappa &=& \beta_\kappa=0.
\end{eqnarray}
The $\beta$ functions for the couplings $\kappa$ and $\lambda_2$ can be
computed from the effective potential flow, yielding
\begin{eqnarray}
\beta_\kappa&=&-2\kappa+\frac{(2N_{\mathrm{L}}-1)}{32\pi^2}
+\frac{3}{32\pi^2(1+2\kappa\lambda_2)^2}-\frac{h^2}{4\pi^2\lambda_2(1+\kappa  h^2)^2},\\ 
\beta_\lambda&=&\frac{(2N_{\mathrm{L}}-1)\lambda_2^2}{16\pi^2}
+\frac{9\lambda_2^2}{16\pi^2(1+2\kappa\lambda_2)^3}-\frac{h^4}{2\pi^2(1+\kappa h^2)^3},
\end{eqnarray}
and the $\beta$ function of the Yukawa coupling reads
\begin{eqnarray}
\beta_h&=&\frac{1}{16\pi^2}\frac{h^4}{(1+\kappa h^2)}\Bigg\{-\frac{6\kappa\lambda_2}{(1+2\kappa\lambda_2)^2}\left( \frac{1}{1+\kappa h^2} + \frac{2}{1+2\kappa\lambda_2}\right)\\
&{}&-\left( \frac{1}{1+\kappa h^2}+ 1\right)+\frac{1}{(1+2\kappa\lambda_2)}\left( \frac{1}{1+\kappa h^2} + \frac{1}{1+2\kappa\lambda_2} \right)\nonumber\\
&{}&+\frac{2\kappa h^2}{(1+\kappa h^2)}\left( \frac{2}{1+\kappa h^2}+1 \right)+2\lambda_2\kappa\left( \frac{1}{1+\kappa h^2} +2\right)\nonumber\\
&{}&-\frac{2\kappa h^2}{(1+\kappa h^2)(1+2\kappa\lambda_2)}\left( \frac{2}{1+\kappa h^2}+\frac{1}{1+2\kappa\lambda_2}\right) \Bigg\}\nonumber\,.
\end{eqnarray}
The explicit derivation of these $\beta$ functions using the Wetterich
equation together with an optimised regulator\cite{Litim:2001up} can be found
in \cite{Scherer:2009cy}.
With these $\beta$ functions we find non-Gau\ss ian fixed points (NGFPs) for
$1\leq N_\mathrm{L} \leq 57$. An extension of the truncation in the effective
potential shows a reliable convergence of the fixed point and its critical
exponents \cite{Scherer:2009cy}. An extension of the fixed-point analysis with
a truncation to next-to-leading order in the derivative expansion by
introducing flowing wave function renormalisations for the left-/right-handed
fermion fields as well as for the boson field is not straightforward, since
the algebraic structure of the $\beta$ functions becomes more involved. We
have found no analytical way to solve those equations. Numerical evidence
suggests that the fixed point might be destabilized by large anomalous
dimensions at next-to-leading order. This is not surprising, as many massless
Goldstone and fermion modes exist in the present model also in the SSB regime
which can induce instabilities. As the standard model does not have these
massless modes, it is natural to expect that this issue at next-to-leading
order is resolved by introducing gauge fields. In
Fig.~\ref{fig:critexp} we show the critical exponents of the basic three
parameter truncations as a function of the left-handed fermion number
$N_\mathrm{L}$.
As an explicit example to show how asymptotic safety leads to predictivity for
toy standard model observables we study a leading-order truncation expanded up
to $\frac{\lambda_6}{6!}\rho^6$ in the effective potential and
$N_\mathrm{L}=10$. We find (non-universal) fixed-point values
\begin{eqnarray}
 &&\kappa^\ast=0.0152, \quad \lambda^\ast=12.13,\quad h^{\ast 2}=57.41.\nonumber
\end{eqnarray}
For the universal critical exponents we obtain
\begin{eqnarray}
 &&\Theta_{1}=1.056,\quad \Theta_2=-0.175,\quad \Theta_3=-2.350.\nonumber
\end{eqnarray}
There is only one relevant direction, corresponding to one physical parameter
to be fixed. All other parameters are predictions of the theory. The exponent
of the relevant direction is 1.056 (as compared to 2 near the Gau\ss ian fixed
point), such that the hierarchy problem is weakened. We will fix the flow by the IR
value of $\kappa$. In a realistic model this would correspond to the vev
(which can be determined from the Z/W-boson masses)
\begin{equation}
v=\lim_{k\to0} \sqrt{2\kappa}k.\nonumber
\end{equation}
The IR values of the other two parameters are predicted by the RG flow starting from the NGFP in the UV and are related to the Higgs and the Top mass.
\begin{equation}
 m_{\mathrm{Higgs}}=\sqrt{ \lambda_2 } v , \quad m_{\mathrm{top}}=\sqrt{ h^2 } v.\nonumber
\end{equation}
Choosing the standard model vev $v=246$GeV the predictions within this toy model are \cite{Scherer:2009cy}
\begin{equation}
 m_{\mathrm{Higgs}}=0.97 v \simeq 239\text{GeV} , \quad m_{\mathrm{top}}=5.78 v\simeq 1422\text{GeV}.\nonumber
\end{equation}

\section{Discussion and Conclusions}

In this contribution we sketched the idea of asymptotic safety and explained
how this scenario could conceptually be applied to solve the problems of
triviality and hierarchy in the standard model of particle physics. Therefore
we use the functional RG in the formulation by Wetterich as a nonperturbative
tool for QFT, and derive flow equations for a chiral Yukawa model with one
right-handed and $N_\mathrm{L}$ left-handed fermions. This asymmetry of the
fermion species allows for a balancing of the fermion and the boson
fluctuations and therefore can generate a NGFP in theory space. We find NGFPs
for $1\leq N_\mathrm{L} \leq 57$ and analyse the properties and predictions of
asymptotic safety explicitely for the example $N_\mathrm{L}=10$. Here we find
a NGFP with one relevant and two irrelevant directions in theory space in a
basic truncation, allowing for a prediction of the toy Higgs and the toy top
mass. 

This result is stable with respect to an extension of the truncation in
the effective potential.  Due to the existence of massless Goldstone and
fermion fluctuations, which are not present in the standard model, we observe
a possible destabilisation at next-to leading order in the derivative
expansion. A more realistic model requires gauge bosons, potentially
stabilizing our scenario. Work in this direction is under way. In fact, the
NGFP discussed in the present work has first been discovered in a simpler
$Z_2$ invariant Yukawa system, where the discrete symmetry does not give rise
to Goldstone bosons in the broken regime \cite{GiesScherer:2009}. Even though
the fixed point exists in the $Z_2$ model only for somewhat esoteric fermion
flavor numbers $\Nf\lesssim 0.3$, the fixed point and its critical properties
have been shown to remain stable also at next-to-leading order in the
derivative expansion.

We conclude that an asymptotically safe gauged version of our model has the
prospect of quantitatively predicting IR observables such as particle masses,
which are free parameters in a perturbative analysis of the standard model. It
would solve the problem of triviality structurally and it could improve the
hierarchy problem. Put on a gravitational background as in \cite{zanusso:2009}
and in connection with the asymptotic safety scenario in quantum gravity this
could constitute a fundamental version of all known interactions which is
valid on arbitrarily large scales.

\section{Acknowledgments}

This work was supported by the DFG under contract No. Gi 328/5-1
(Heisenberg program), FOR 723 and GK1523/1. M.M.S. is grateful to the organizers of the 49th Cracow School of Theoretical
Physics for the opportunity to present this work.

\end{document}